%% file: main.tex
\documentclass{IEEEcsmag}
\usepackage[colorlinks,urlcolor=blue,linkcolor=blue,citecolor=blue]{hyperref}

\usepackage{upmath}
\usepackage{graphicx} 
\usepackage{caption,subcaption}

\usepackage{hyperref}

\newcommand{\revision}[1]{\textcolor{black}{#1}}
\jvol{XX}
\jnum{XX}
\paper{XX}
\jmonth{X/X}
\jname{IEEE Computer Graphics and Applications}
\pubyear{2020}

\begin{document}

\sptitle{Department: Applications}
\editor{Editor: Mike Potel, potel@wildcrest.com}

\title{Cartographic Design of Cultural Maps}

\author{\ Edyta Paulina Bogucka}
\affil{\ Technical University of Munich}

\author{\ Marios Constantinides}
\affil{\ Nokia Bell Labs, Cambridge (UK)}       

\author{\ Luca Maria Aiello}
\affil{\ Nokia Bell Labs, Cambridge (UK)}

\author{\ Daniele Quercia}
\affil{\ Nokia Bell Labs, Cambridge (UK)}

\author{\ Wonyoung So}
\affil{\ Massachusetts Institute of Technology}

\author{\ Melanie Bancilhon}
\affil{\ Washington University in Saint Louis}

\markboth{Cultural Maps}{Cultural Maps}

\begin{abstract}
Throughout history, maps have been used as a tool to explore cities. They visualize a city's urban fabric through its streets, buildings, and points of interest. Besides purely navigation purposes, street names also reflect a city's culture through its commemorative practices. Therefore, cultural maps that unveil socio-cultural characteristics encoded in street names could potentially raise citizens' historical awareness. But designing effective cultural maps is challenging, not only due to data scarcity but also due to the lack of effective approaches to engage citizens with data exploration. To address these challenges, we collected a dataset of 5,000 streets across the cities of Paris, Vienna, London, and New York, and built their cultural maps grounded on cartographic storytelling techniques. Through data exploration scenarios, we demonstrated how cultural maps engage users and allow them to discover distinct patterns in the ways these cities are gender-biased, celebrate various professions, and embrace foreign cultures.
\end{abstract}

\maketitle
\begin{IEEEkeywords}
H.2.8.c Data and knowledge visualization; O.6.1 Methods of data collection; I.2.1.a Cartography
\end{IEEEkeywords}

\input{sections/1_introduction}

\input{sections/2_dataset}
\input{sections/3_viz}

\input{sections/4_discussion}

\input{references}

\end{document}

%% file: sections/1_introduction.tex
\chapterinitial{Introduction} 

The European Council defines intercultural cities as urban units with diverse population from different nationalities, origins, languages, or religions and beliefs~\cite{intercultural_cities}. It crafted strategic actions for reviewing current urban policies of cultural inclusion. This agenda prompts cities to collect data and informal stories on communities, develop participatory methods of citizens' engagement, and lead public awareness campaigns~\cite{intercultural_cities}. 

Various data collection and visualization techniques can be employed to explore cultural diversity~\cite{cultural_analytics}. Among them \emph{maps} are widely used to make spatial patterns of cultural production visible. The earliest examples of cultural maps are archival administrative map series, which document historical place naming processes~\cite{streets_poland_czech}. Due to the developments in web mapping, maps have become social justice tools promoting indigenous territories (\url{https://native-land.ca}), exposing cultural diversity indices~\cite{cultural_mapping2015}, or giving voice to marginalized groups. For example, the Southern Poverty Law Center mapped the public oppression symbols of the Civil War still present in American street names (\url{https://www.splcenter.org/data-projects/whose-heritage}), while the feminist association l'Escouade developed an alternative map of Geneva to celebrate female influence through a symbolic renaming of 100 streets (\url{https://100elles.ch}). To assess cultural diversity, cities need to first develop mechanisms to quantify their culture; a challenging task due to official administrative data scarcity. Therefore, researchers often resort to unconventional datasets to develop proxies that capture complex socio-cultural characteristics. These include geo-referenced pictures~\cite{hristova2018}, points of interests ~\cite{social_media}, social media photos~\cite{cultural_analytics}, and street names~\cite{streets_city_text}, to name a few. 

Associating a city's street naming system with socio-cultural dimensions emerged in the 18\textsuperscript{th} century, when street naming practices switched from highlighting cities' geographic features to establishing commemorative spaces~\cite{streets_city_text}. Street names were used to develop urban indicators of male predominance, women's societal role, religious connections~\cite{streets_city_text}, and biodiversity~\cite{streets_biodiversity}. What remains less explored, however, are comparative studies of cities. Examples in this direction are the discovery of commemorative practices of new settlements in Israel~\cite{streets_israel}, socialistic Poland and Czechia~\cite{streets_poland_czech}. But case-based research lacks automated workflows and is mostly done manually by mapping communities or by participants of thematic workshops. For example, the Open Street Maps (OSM) mapping community Geochicas made it possible to investigate the gender distribution of street honorees of eleven Spanish speaking cities including Barcelona, Havana, and Mexico City (\url{https://geochicasosm.github.io/lascallesdelasmujeres}). Overall, previous street names studies have been limited in scale (covering one city or group of cities in the same country), have been restricted to a single cultural dimension, or have lacked appealing visual interfaces to raise awareness about the commemorative practices. 

While maps have been widely used as tools for cities' exploration, they have not yet been adopted as visual analytical tools to reason about cities' culture and create awareness of ``who'' stands behind a certain street~\cite{streets_poland_czech}. To increase citizens' engagement and awareness, one needs to ensure users' emotional responses~\cite{citizen_engagement}. One way to achieve that, cartographers say, is through visual storytelling~\cite{roth_2020}, which aims at evoking surprise~\cite{emotional_appeals_2016} and inducing reflection~\cite{style_effect_2012}. Alternatives include interactive functionalities that incorporate the audience's voice or show the humans behind the data~\cite{napa_cards}. Current street names maps use simple thematic mapping techniques to color-code streets based on the honoree's mother language (\url{http://str.sg/sgstreets}),  gender (\url{http://genderatlas.at}), (\url{https://equalstreetnames.brussels}), or occupation (\url{https://www.zeit.de/feature/streetdirectory-streetnames-origin-germany-infographic-english}). Therefore, the main challenge of cultural mapping that remains is how to develop a visual language that represents intangible values encoded in the data such as a city's global focus or gender bias. To help meet this challenge, our work makes two main contributions: 
    \begin{itemize}
    \item a workflow for street names collection in the cities of Paris, Vienna, New York, and London (\S\nameref{sec:section3}). 
    \item interactive cultural maps, grounded on cartographic techniques, which promote historic awareness while doing so in a playful way (\S\nameref{sec:section4}).
    \end{itemize}

%% file: sections/2_dataset.tex
\section{DATASET}
\label{sec:section3}

From open data sources, we collected and curated a dataset of 4,932 streets in four cities, namely Paris, Vienna, New York, and London, which we describe next. For each street, we gathered eight types of information (Table~\ref{tab:dataset}). 

\subsubsection{Paris}
Using the SPARQL endpoint query service of Wikidata (\url{https://query.wikidata.org}), we retrieved a total of 3,413 streets located in Paris. Each returned street object, including the field \emph{named after}, which corresponds to an eponym (e.g., event or person). We used it to filter streets that are named after a person by  only taking the eponym instances of the \emph{person class}. This yielded a dataset of 1,808 streets. We then crawled these persons' Wikipedia page and obtained each street's: borough, honoree's name, gender, occupation, date of birth, date of death, and country of origin. After further data cleaning, the final Paris dataset contains information for 1,428 honorific streets, and covers the historical period between 1202 and 2011. 

\subsubsection{Vienna}
The `Vienna History Wiki' platform (\url{https://www.geschichtewiki.wien.gv.at}) aggregates historical knowledge about the city in a structured way (similar to Wikipedia). We crawled the webpages containing the history of a given street, and retrieved an initial dataset of 2,481 streets, along with the eight type of information for each street. All German information was translated to English using a Python \emph{language translate} package. After data cleaning, the final Vienna dataset contains 1,662 streets, and covers the historical period between 1778 and 2018.

\subsubsection{London}
As no curated London dataset was publicly available, we resorted to Wikipedia entries that referred to London streets. We narrowed down the search space using mentions that typically characterize pages about streets such as `named after', `honor', and `celebrate'. This resulted into a dataset of 2500 streets, which were annotated along the eight types of information. Each street was annotated by at least 2 people through a crowdsourcing task using the Amazon Mechanical Turk platform, and we manually resolved conflicts. After data cleaning, the resulting London dataset contains 770 streets, and covers the historical period between 1030 and 2013.
 
\subsubsection{New York}
We obtained a curated dataset of 1,459 honoric streets from the urban planner Gilbert Tauber (\url{http://nycstreets.info}). The dataset is aligned with the eight types of information for each street. After data cleaning, the New York dataset contains 1,072 streets, and covers the historical period between 1998 and 2013.

\begin{table}
\caption{Dataset structure. Each row corresponds to a street for which name, location, and honoree information have been collected.}
\label{tab:dataset}
\small
\begin{tabular*}{17.5pc}{@{}|p{45pt}<{\raggedright}|p{140pt}<{\raggedright}|@{}}
\hline
Field& 
Description\\
\hline
streetname & Name of the honorific street \\
district & District that the street belongs to \\
denomination & (Re)naming date of the street \\
honoree & Person that the street was named after \\
gender & Gender of the person \\
occupation & Occupation of the person \\
country & Country of origin of the person \\
dob & Date of birth of the person \\
dod & Date of death of the person \\
\hline
\end{tabular*}
\end{table}

\subsubsection{Dataset preparation}
To ensure consistency across the four datasets, we utilized a unified coding scheme for occupations. Based on the International Standard Classification of Occupations, we mapped each occupation to any of the following 17 occupational groups: legislators, writers, creative and performing artists, science and engineering professionals, health associate professionals, sportsmen, social workers, teaching professionals, businessmen, craft and related trades workers, legal and social professionals, religion representatives, military personnel, royals, politicians, and 9-11 Responders and Victims. 
Additionally, we matched each street with its corresponding OSM shape file. To build the cultural maps, we stored the underlying geodata in the PostgreSQL database, and designed the interface with Mapbox GL JS library (\url{https://docs.mapbox.com/mapbox-gl-js/api}).

\begin{figure}
    \centering
    \includegraphics[width=0.95\linewidth]{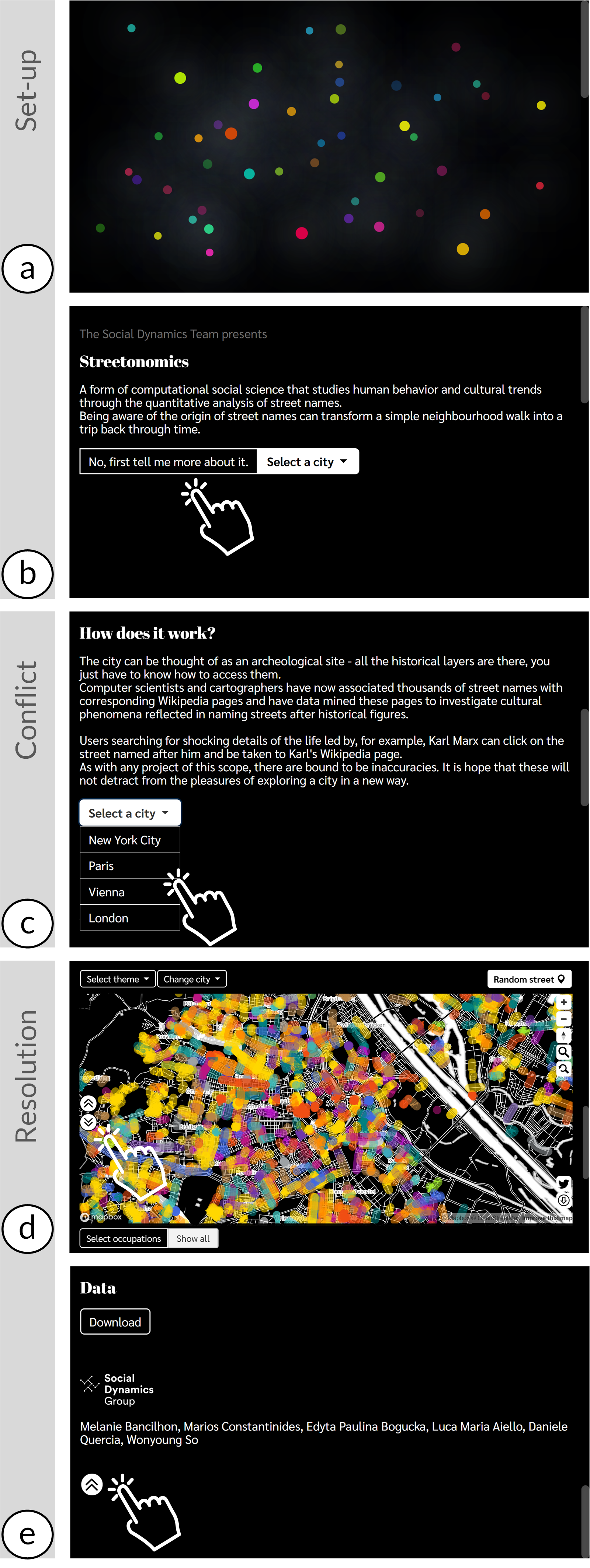}
    \caption{Narrative design of cultural maps.
    (a) Setting a visual tone with the animation of flying points. (b) A promise of an unusual neighbourhood walk serves as a hook to capture audience attention. (c) The problem of street names data as cultural indicators. (d) Cultural maps make street naming patterns visible. (e) Final denouement with data, credit sections, and reading materials. User advances the plot through entry points marked with the click icon. \revision{Video available: \url{http://social-dynamics.net/streetonomics/teaser.mp4}}}
    \label{fig:storyflow}
\end{figure}

%% file: sections/3_viz.tex
\section{CULTURAL STREET MAPS}

Next we introduce the storytelling techniques used to develop cultural street maps, explain their user interface, and provide map exploration possibilities through a narrative walk-through.

\begin{figure}
    \centering
    \includegraphics[width=\linewidth]{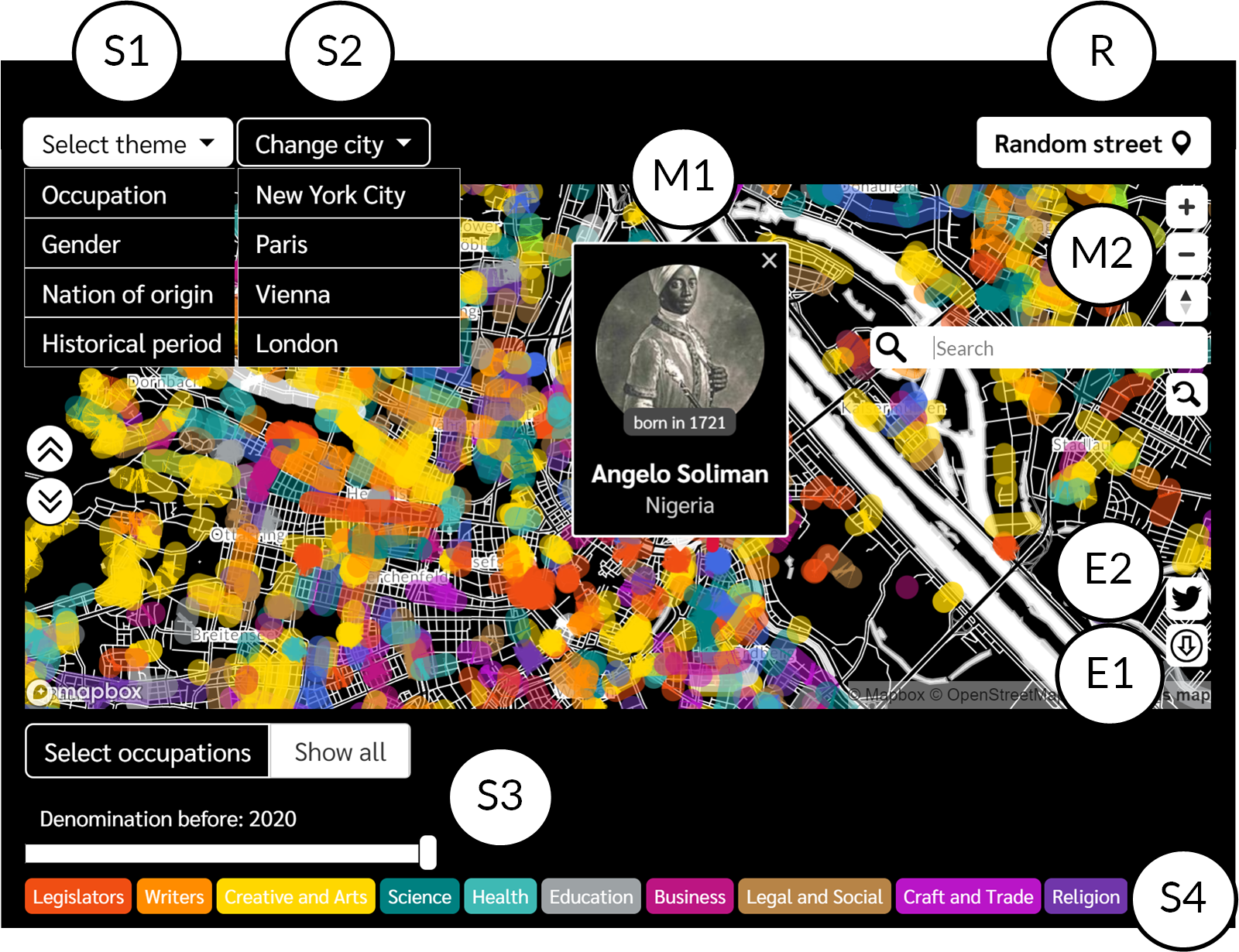}
    \caption{Four functional regions of the user interface: group selections (S1-S4), map controls (M1-M2), serendipitous discoveries (R), and user engagement functionalities (E1-E2).}
    \label{fig:ui_regions}
\end{figure}

\begin{figure*}
\begin{subfigure}{0.45\textwidth}
\centering 
  \includegraphics[width=\linewidth]{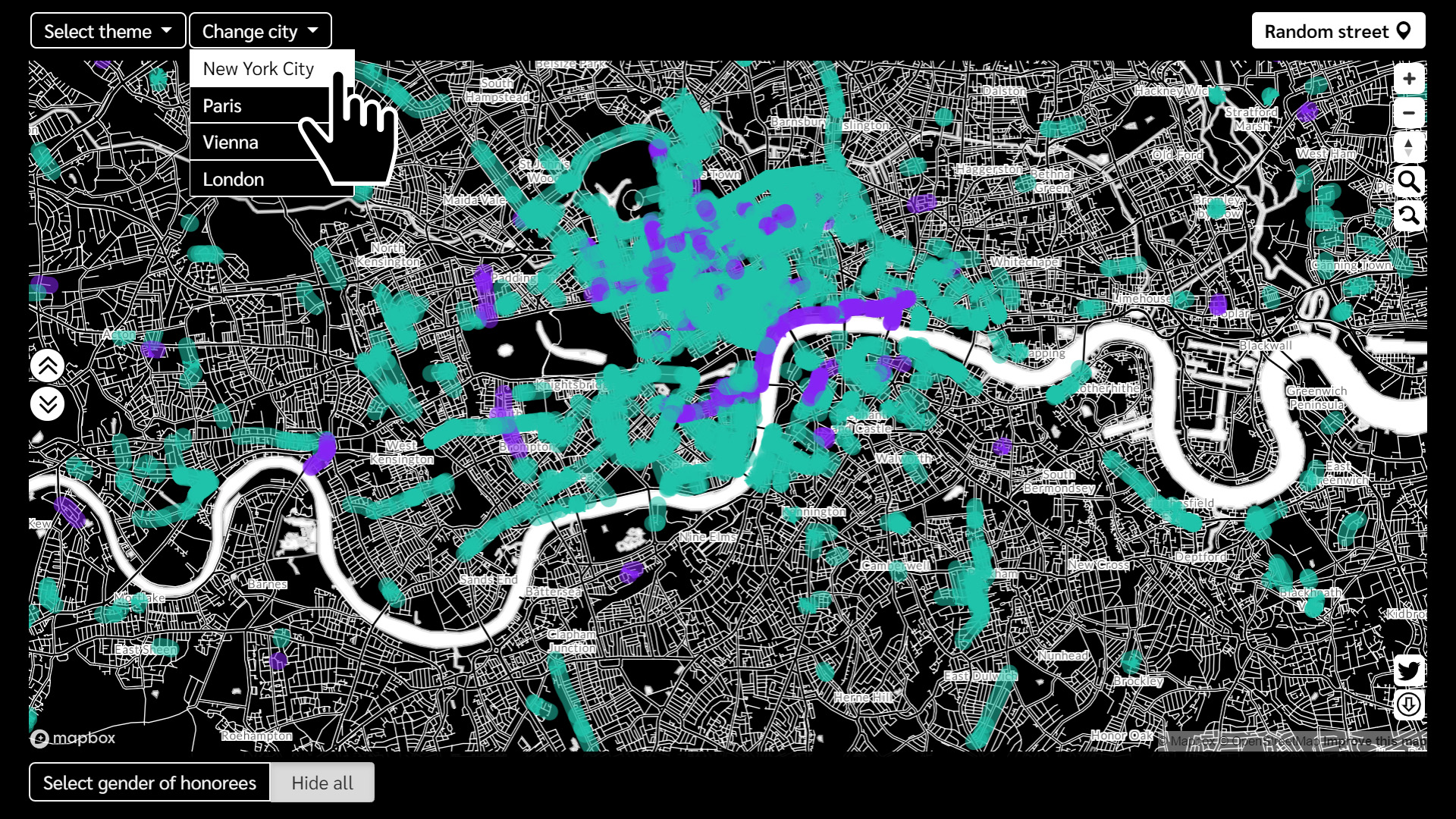}
  \caption{S1: selecting a city of interest}
  \label{fig:task_changeCity}
\end{subfigure}\hfil 
\begin{subfigure}{0.45\textwidth}
\centering 
  \includegraphics[width=\linewidth]{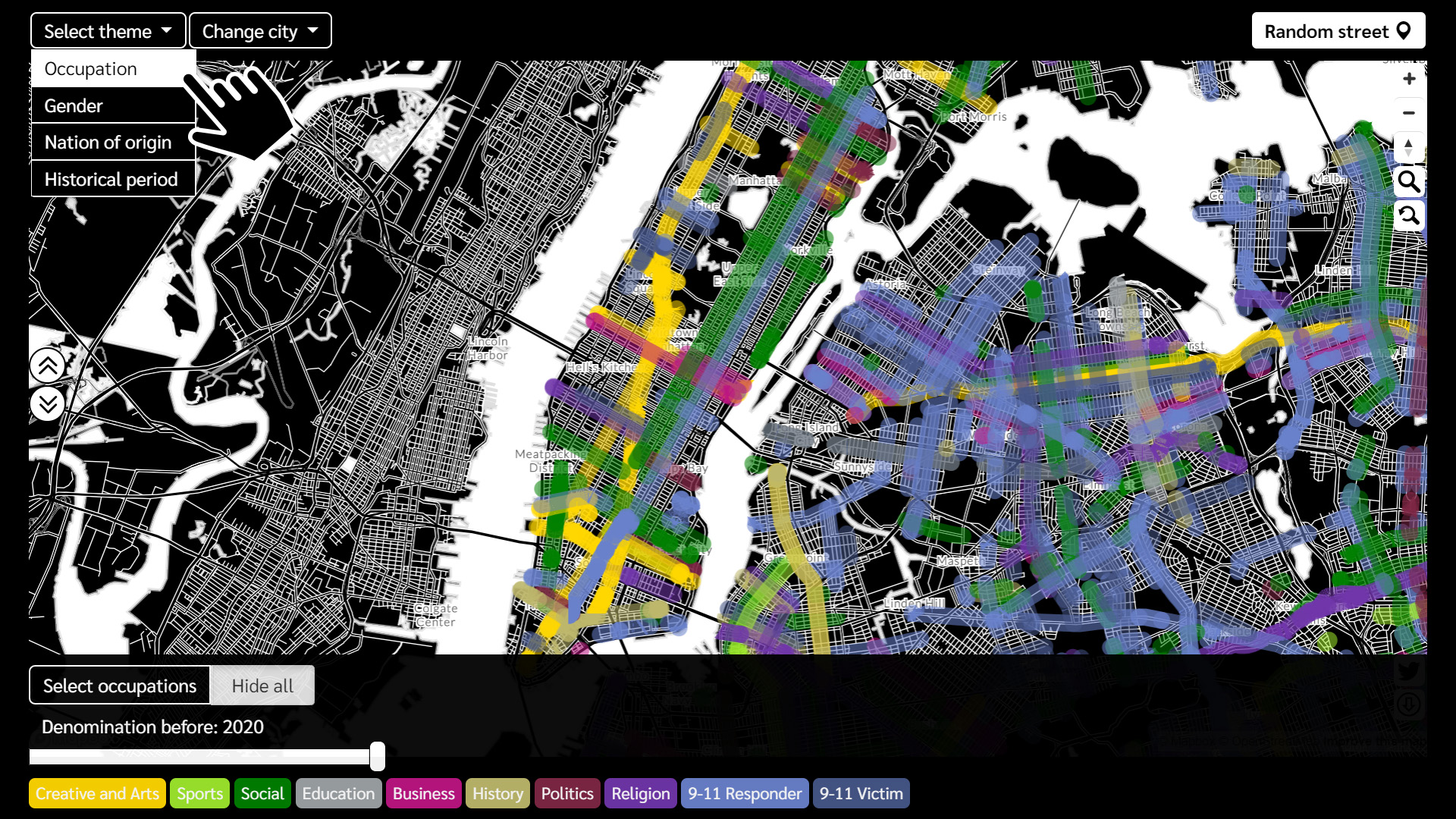}
  \caption{S2: changing the map’s theme}
  \label{fig:task_SelectTheme}
\end{subfigure}\hfil 

\medskip
\begin{subfigure}{0.45\textwidth}
\centering 
  \includegraphics[width=\linewidth]{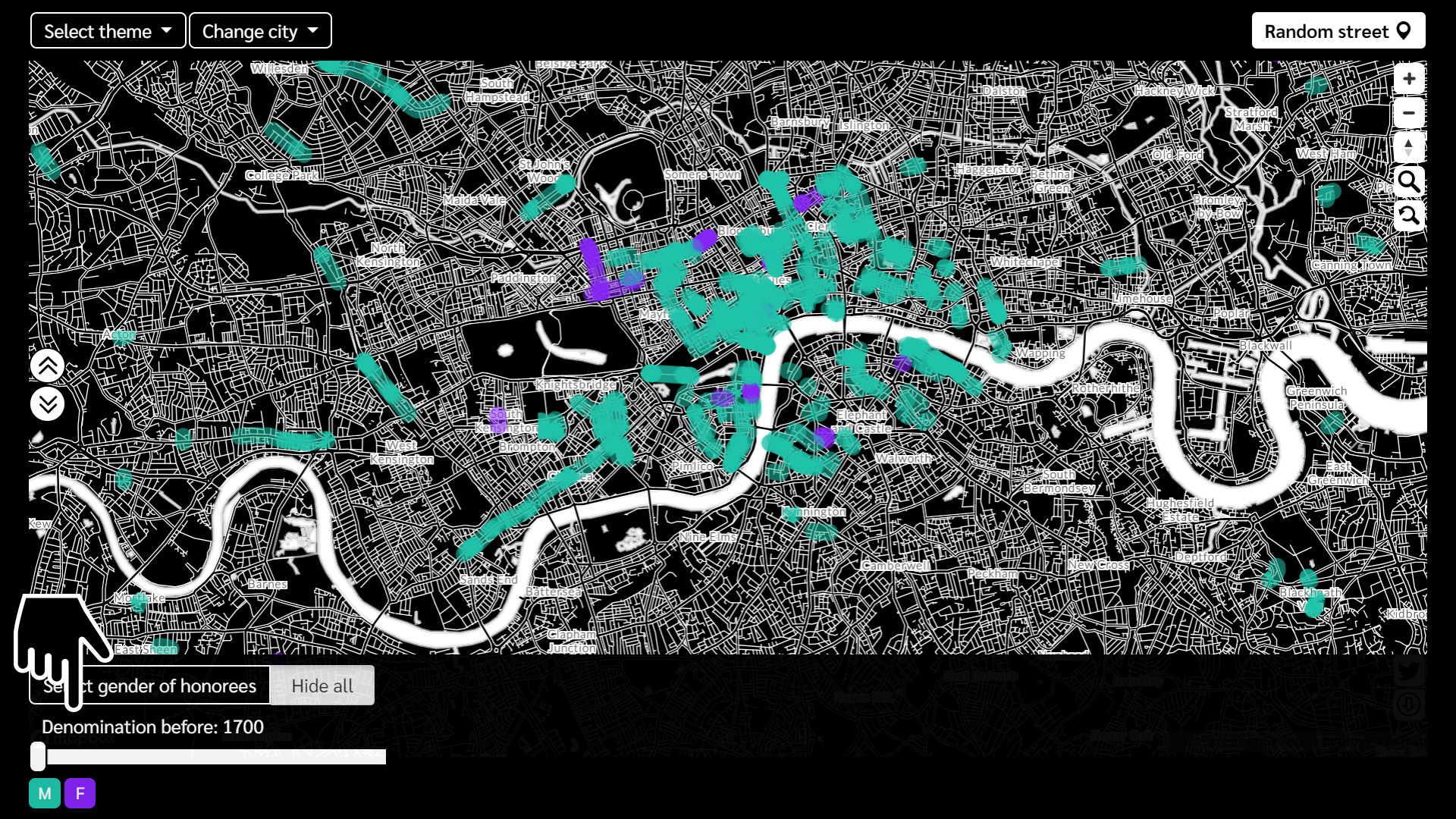}
 \caption{S3: visualizing streets (re)named in a given period}
  \label{fig:task_category}
\end{subfigure}\hfil
\begin{subfigure}{0.45\textwidth}
\centering 
  \includegraphics[width=\linewidth]{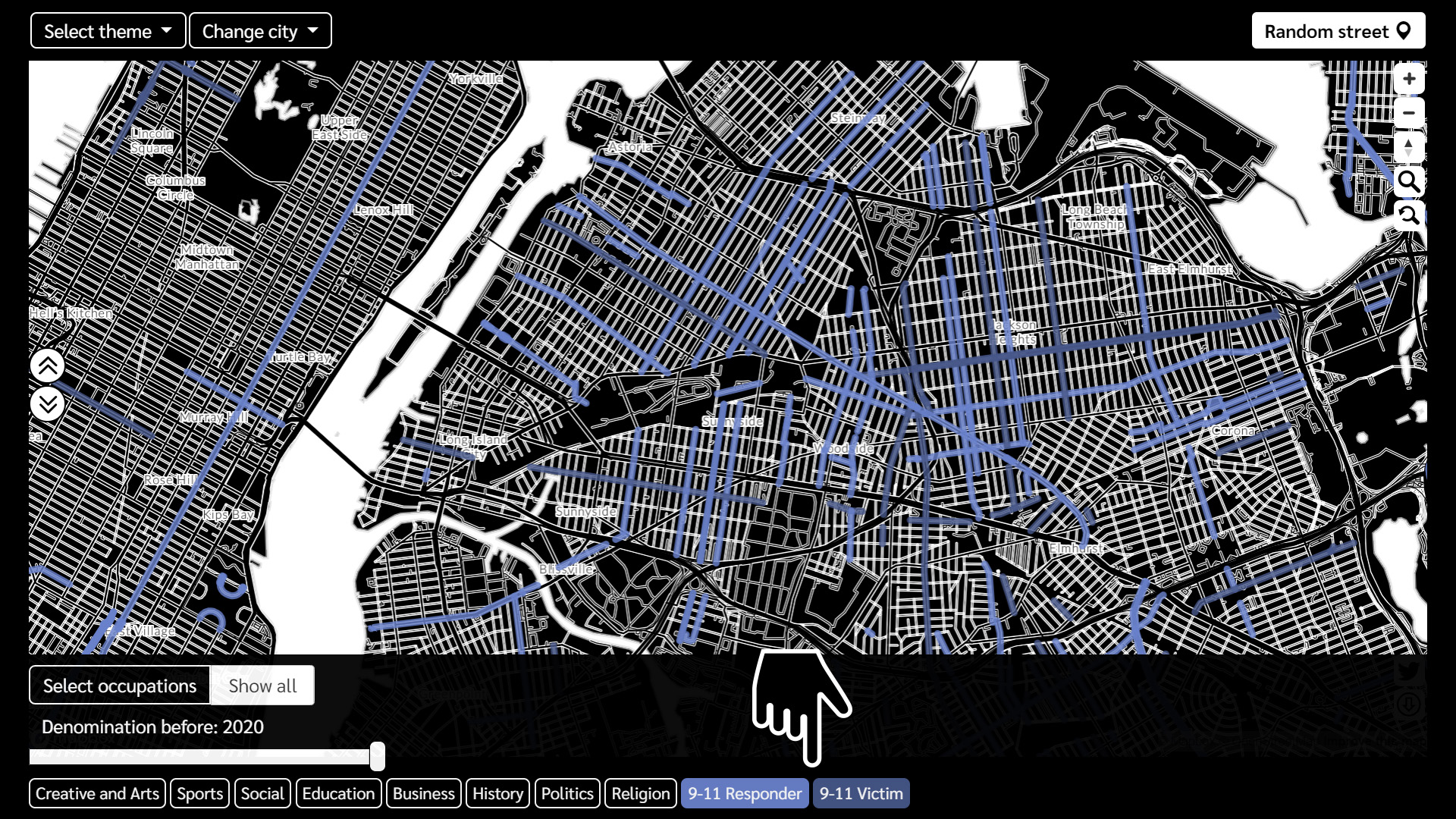}
 \caption{S4: visualizing streets categories by honorees}
  \label{fig:task_denomination}
\end{subfigure}\hfil

\medskip
\begin{subfigure}{0.45\textwidth}
\centering 
  \includegraphics[width=\linewidth]{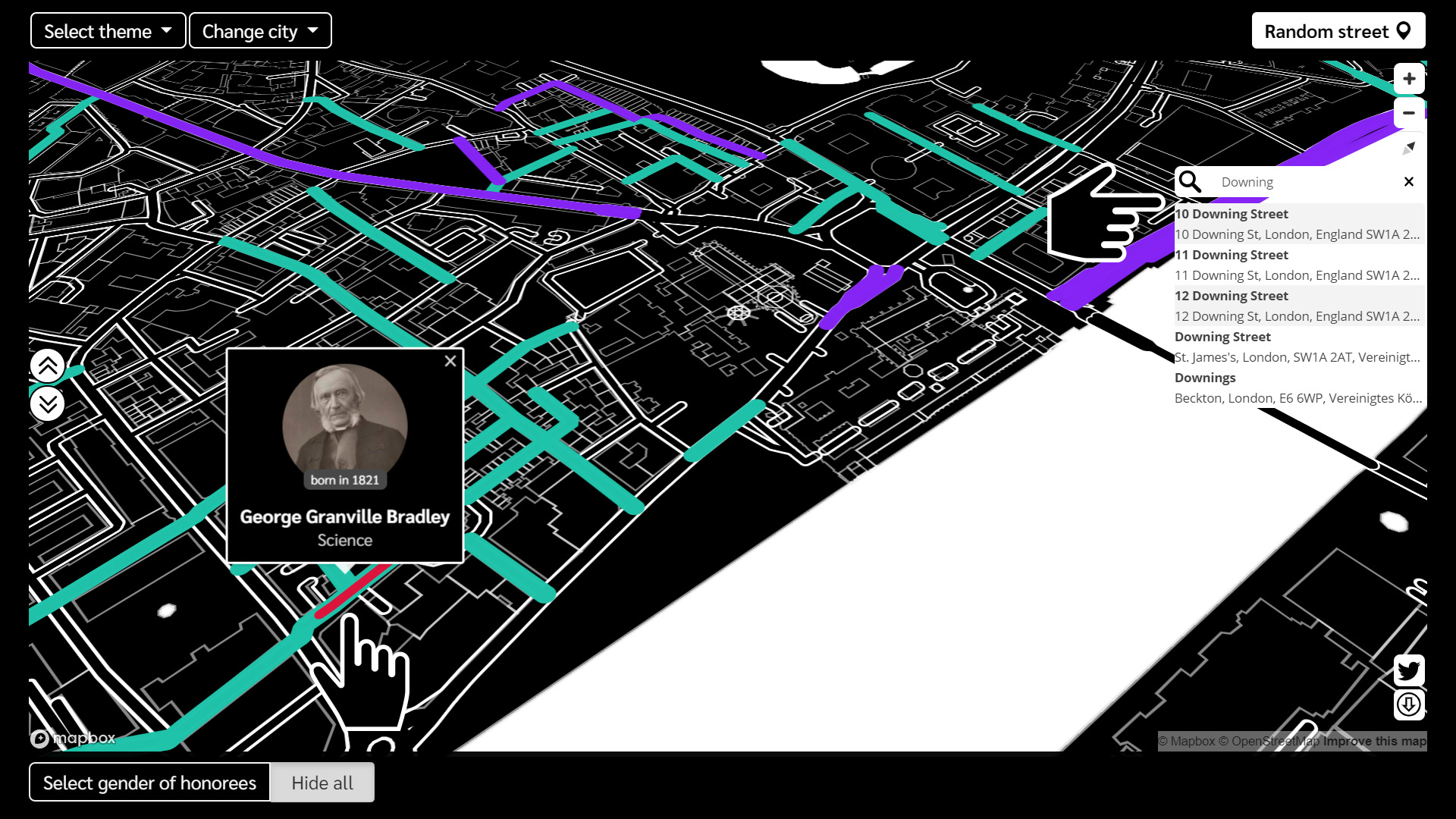}
 \caption{M1, M2: exploring the map in-depth}
  \label{fig:task_CasualExploration}
\end{subfigure}\hfil
\begin{subfigure}{0.45\textwidth}
\centering 
    \includegraphics[width=\linewidth]{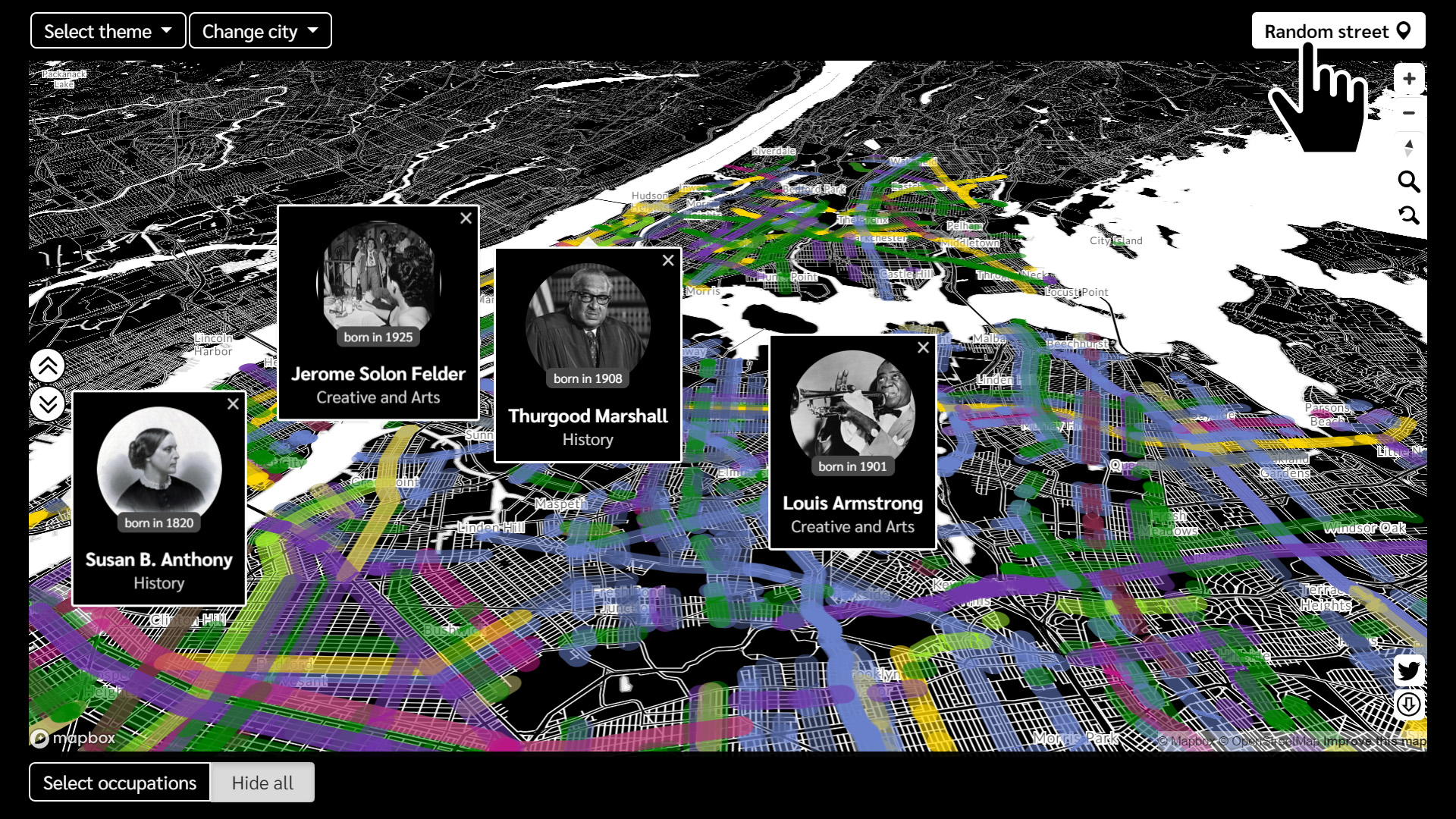}
 \caption{R: focusing on a random street}
 \label{fig:task_randomButton}
\end{subfigure}\hfil

\medskip
\begin{subfigure}{0.45\textwidth}
\centering 
  \includegraphics[width=\linewidth]{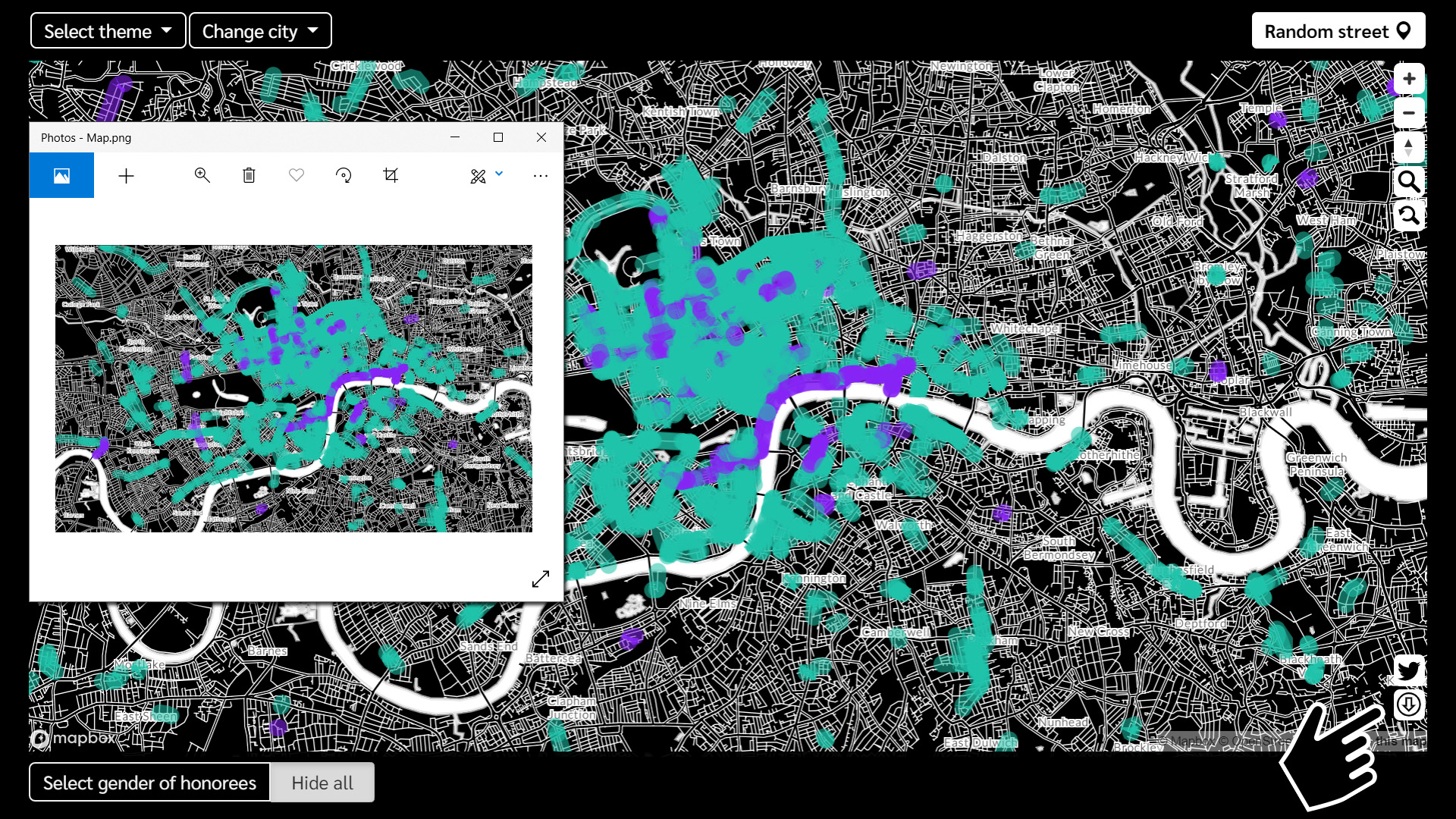}
 \caption{E1: downloading a picture of the map}
  \label{fig:task_screenshot}
\end{subfigure}\hfil
\begin{subfigure}{0.45\textwidth}
\centering 
  \includegraphics[width=\linewidth]{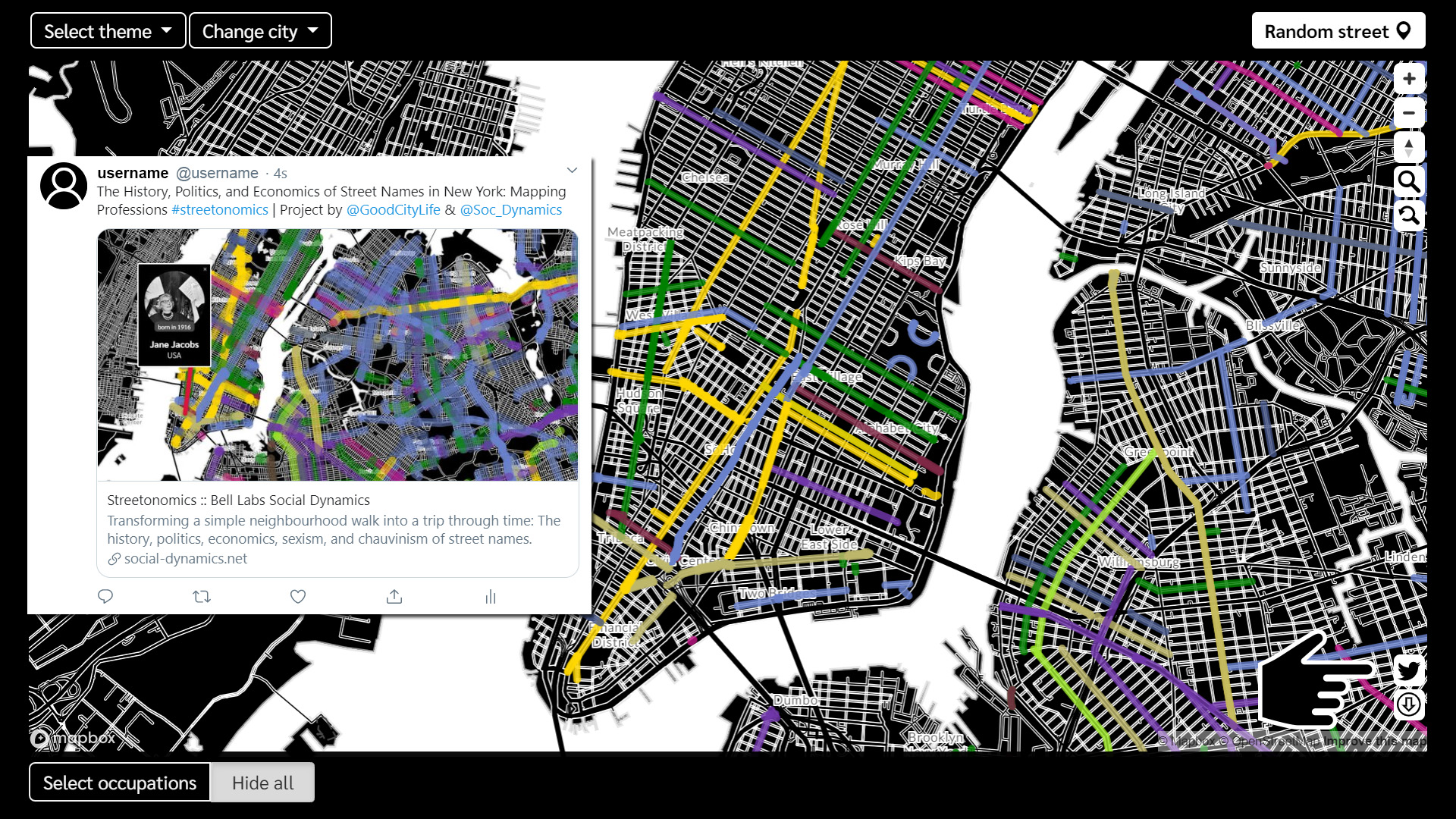}
 \caption{E2: sharing insights on social media}
  \label{fig:task_tweet}
\end{subfigure}\hfil
\caption{Interaction capabilities of cultural maps involve selections \emph{(a,b,c,d)}, map controls \emph{(e)}, button for  serendipitous discoveries \emph{(f)}, and user engagement buttons \emph{(g,h)}.}
\label{fig:tasks}
\end{figure*}

\label{sec:section4}
\subsection{Cartographic techniques for visual storytelling}

Cultural street maps are developed using a map-based visual storytelling conceptual framework~\cite{roth_2020}, \revision{and are publicly accessible (\url{http://social-dynamics.net/streetonomics})}. They are organized into a \emph{spatial narrative} (plotline) and utilize \emph{tropes} to introduce the story.

The visualization consists of five sections, ordered in a three-arc narrative sequence from set-up through conflict to resolution (Figure~\ref{fig:storyflow}), to show street names patterns in each city and between cities. The first section in the \emph{set up} contains a pre-loader with a micro animation to indicate the visual tone of the application and create suspense (Figure~\ref{fig:storyflow}a). The second section captures audience's attention by a \emph{hook}; a promise of an unusual neighborhood walk (Figure~\ref{fig:storyflow}b). The third section introduces \emph{conflict} and presents the \emph{problem context} of street names as cultural indicators (Figure~\ref{fig:storyflow}c). 
The fourth one presents cultural maps as the \emph{resolution} (Figure~\ref{fig:storyflow}d). The visualization makes the \emph{problem context} visible and encourages users to explore a city. The last section includes a \emph{denouement} with original data, attributions, and additional materials (Figure~\ref{fig:storyflow}e). 

To engage users in the exploration of cultural maps, we used six visual storytelling tropes:

\begin{enumerate}

\item \emph{Pointillism} is a \emph{metaphorical} impressionistic technique which applies paint through colorful spots~\cite{pointilism2008}. When such image is viewed from a distance, dots are blended by the viewer’s eye into a comprehensible spatial pattern. Similarly, overlapping points of single, seemingly unconnected streets form commemorative landscapes (Figure~\ref{fig:storyflow}d) and keep the \emph{mood} of suspense.

\item \emph{Zoomy-telling} is our name for a visual narration technique that adjusts visualization content depending on the selected zoom level. Cultural maps offer different representations of data on the view levels of the city, district and street. While looking at a city as whole, a user sees colorful, painting-like thematic data clusters (Figure~\ref{fig:storyflow}d, Figure~\ref{fig:tasks}c). When zoomed to district level, streets turn into half-transparent lines (Figure~\ref{fig:tasks}d) and then become fully opaque at street level (Figure~\ref{fig:tasks}e).

\begin{figure*}
    \centering
    \includegraphics[width=\linewidth]{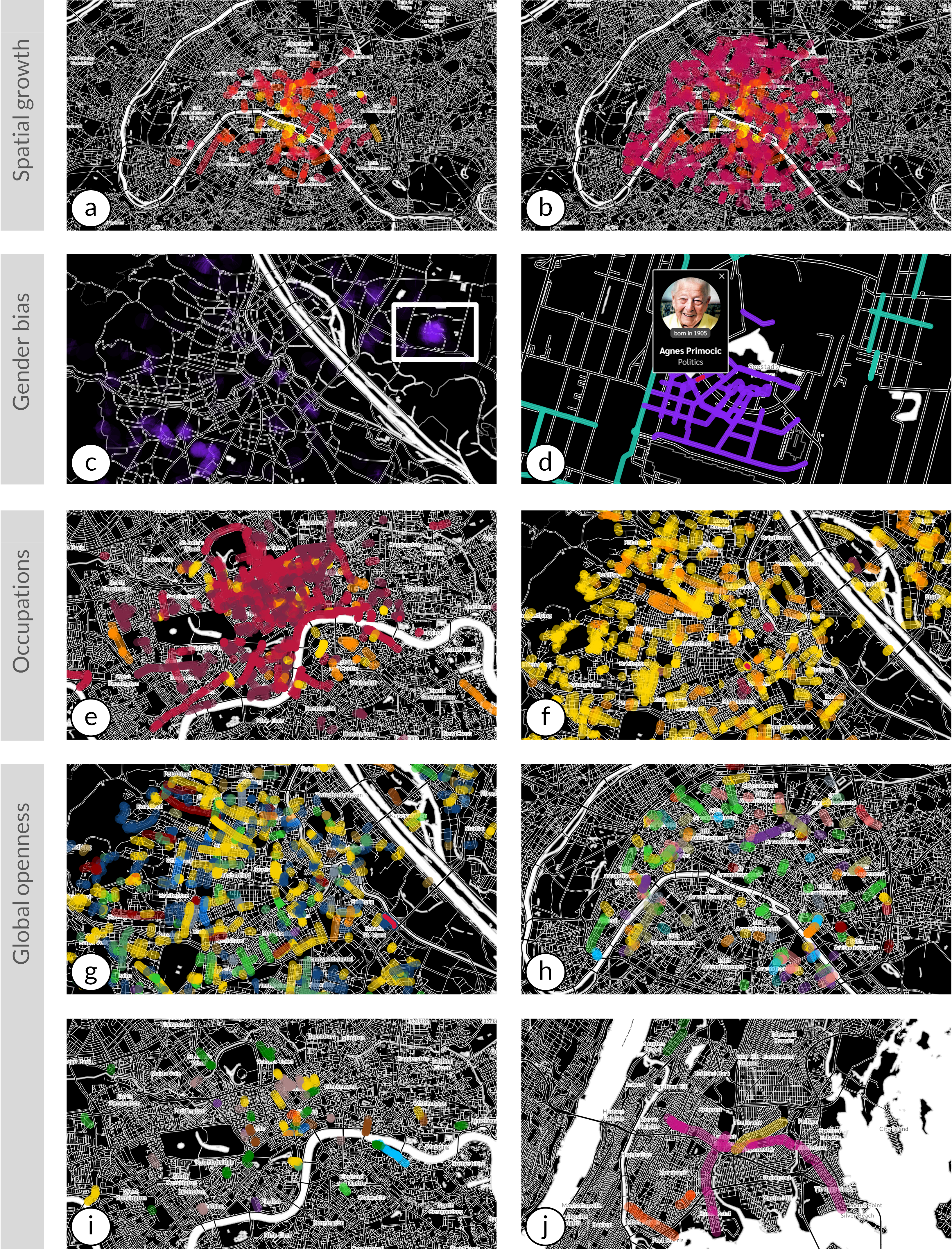}
    \caption{Scenarios of map-based explorations: Urban growth of Paris during the Hausmann's enlargement phase between 1853 (a) and 1870 (b).  
    Cluster of streets named after females near Seestadt Aspen (c) reflects recent Viennese naming policies (d). 
    Celebrating the popularity of Royals and politicians (marked in red tones) in London (e) vs. paying tribute to artists and writers (marked in yellow tones) in Vienna (f). Vienna is open to a variety of nationalities (g), while Paris (h), London (i) and New York (j) are less so.}
    \label{fig:walkthrough}
\end{figure*}

\item \emph{Attention} is influenced by the use of visual hierarchy and highly contrasted figure-ground background map. The level of detail is minimal, with district names being the only labels available for the city's orientation. Each thematic layer (occupation, gender, nation of origin, historical period) is represented using bright, vibrant colors. In doing so, similar occupations are depicted with similar color hues (e.g., streets named after creatives are displayed in tones of orange, while greens indicate professionals from social and science domains). Contrary to the conventional and stereotypical pink and blue color scheme for gender, we resorted to purples and greens as advocated by the data visualization community~\cite{data_feminism}.
Colors of the categorical labels of origin were chosen based on the countries' flags respectively. Finally, the historical period layer utilizes a multi-hue sequential color scheme to map the oldest streets in yellow and the newest ones in blue tones.

\item In spatial narratives, both places and people can be treated as \emph{story characters} ~\cite{roth_2020}, and that allows users to see that each street represents a unique historical persona. Therefore, each time a user selects a street, a styled map pop-up is displayed. The design of the pop-up takes the form of a badge, and contains an image of the honoree, his or her date of birth, name, and profession.

\item The users can use the visualization to express their own \emph{voice} about street names practices by downloading the maps, commenting and sharing them in social media.
\end{enumerate}

\subsection{User Interface and Interactions}
Adhering to responsive design principles, the user interface is organized in four functional regions (Figure~\ref{fig:ui_regions}): selections (S), map controls (M), serendipitous discoveries (R), and user engagement functionalities (E). These serve the following purposes:
 
\begin{itemize}
\item Four selectors (S1-4) promote the exploration of individual streets, their different occupational categories and other themes, and their denomination years.
\item Two types of map controls (M1-2) support transitions between different levels of detail.
\item One random street button (R) provides an opportunity for serendipitous discoveries.
\item Two engagement buttons (E1-2) allow for map download and social media sharing.
\end{itemize}

Cultural maps facilitate an in-depth and visual exploratory analysis. In particular, a user can:
% (Figure~\ref{fig:tasks})
\begin{description}
\item[S1] Select a city of interest (Figure~\ref{fig:tasks}a). 
\item[S2] Select a specific attribute (theme) of the data to be mapped such as `gender' or `occupation' (Figure~\ref{fig:tasks}b). 
\item[S3] Select a given period of time and display subset of streets that were re(named) during that period (Figure~\ref{fig:tasks}c).
\item[S4] Select theme tags to filter and visualize only a subset of streets that were named (or renamed) after a specific group of people. For example, a user can select tags for writers under the theme of `occupation' (Figure~\ref{fig:tasks}d). 

\item[M1] Click on a street segment to enable a pop-up window containing information about the street honoree (including a Wikipedia's link) (Figure~\ref{fig:tasks}e). 
\item[M2] Explore the map by zooming in/out, rotating it, searching for a specific address, and resetting the current view. 

\item[R] Serendipitous exploration by pressing the ``Random street'' button: the map zooms to a randomly selected street and displays a popup window (Figure~\ref{fig:tasks}f). 

\item[E1] Download the currently displayed cultural map as a picture (Figure~\ref{fig:tasks}g). 
\item[E2] Share a cultural map on social media (Figure~\ref{fig:tasks}h). 
\end{description}

\subsection{A narrative walk-through}
By combining nine types of interactions users are encouraged to develop their own data-driven questions. Next we describe the possible scenarios for a single theme and city exploration as well as comparative analysis of cities:

\begin{itemize}

\item \emph{Exploration of a city's spatial development}.
Denomination dates of streets might suggest urban growth. To visualize it, a user can select the city of interest (S1) and map the theme of `historical period' (S2). By dragging the time slider (S3), a mesmerizing animation is revealed. This illustrates the street network growth over the decades and centuries. As an example, consider the effects of Haussmann's enlargement plan for Paris between 1853 and 1870 (Figure~\ref{fig:walkthrough}a-b). %\revision{While keeping the time filter, a user can change the city (S1) to London and see that London reached a similar city area much earlier, with its street names assigned well before the 18\textsuperscript{th} century. Further changing the cities and comparing the dominant street colors helps to place naming processes on timeline, as each city is associated with one color - from the oldest `yellow' London, through `red' Paris, `purple' Vienna to `blue' New York.}
\revision{Using the same time filter, a user can switch (S1) to London and see that the British capital reached a similar city area well before the 18\textsuperscript{th} century. By further switching cities and comparing the dominant street colors helps users to place naming processes on the timeline. This is achieved using one color, starting from the oldest `yellow' London, to the `red' Paris, to the `purple' Vienna to the newest `blue' New York.}

\item \emph{Exploration of historical gender biases and modern responses to mitigate it}. This exploration is available by selecting `gender' as the map's theme (S2), selecting the city of interest (S1), and then the streets (re)named within a specific time period and gender (S3, S4). For example, using the map exploration tools (M1, M2), one could find that street naming policies in Vienna led to the emergence of clusters of streets named after female figures in the last decade (Figure~\ref{fig:walkthrough}c-d). \revision{ Similar newly-formed clusters are present in south Paris (13\textsuperscript{th} arrondissement) and New York (Brooklyn).}
% southern parts of Paris
\item \emph{Exploration of the rise and fall of occupations throughout history} 
User selects the city of interest (S1) and `occupation' as the map's theme (S2). By toggling tags with specific professions (S4) one could see that numerous London streets celebrate the members of the British royal family and politicians, but only a few celebrate artists and writers (Figure~\ref{fig:walkthrough}e). In the next step, one could download the resulting map (E1), change the city to Vienna (S1), and observe that those two cities have opposite commemorative practices (Figure~\ref{fig:walkthrough}f). \revision{ Additionally, one can observe that New York celebrates social activists and no military professionals, while Paris commemorates generals and soldiers from French Revolution, Napoleonic wars and both World Wars.}

\item \emph{Exploration of cities' global openness} starts with selecting `nation of origin' as the map's theme (S2). It continues by choosing multiple cities of interest (S1), downloading their thematic maps (E1), and comparing the patterns. Figure~\ref{fig:walkthrough} shows the country of origin of the foreign street name honorees in Vienna, Paris, London and New York. These maps confirm that, while the commemorative landscape of Vienna reflects many nationalities \revision{from the times of Austro-Hungarian Empire} (Figure~\ref{fig:walkthrough}g), other cities are less open to celebrating foreigners (Figure~\ref{fig:walkthrough}h-j). One could set the downloaded maps (E1) side-by-side and share these insights on social media (E2).

\end{itemize}

%% file: sections/4_discussion.tex
\section{CONCLUSION}
\label{sec:section7}

Cultural street maps visualize a city's history encoded in its streets names through cartographic lenses. Unlike previous interactive street names maps, our approach offers a cartographic perspective in a three-arc narrative structure to promote commemorative practices. Using map-based tropes such as pointillism, zoomy-telling, attention, mood, and voice, it enables reflection of historical processes and keeps users engaged in exploration. In the intercultural cities context~\cite{intercultural_cities}, our tool can encourage the general public to explore hidden city patterns and promote historic awareness. Additionally, it can assist policy-makers to reflect on a city's past and steer its future development. For example, participants of a mapping workshop in Brussels proposed a list of future street honorees to better reflect the city's cultural diversity. Our tool shows that it is possible to advance such initiatives in a productive yet playful way.

%% file: references.tex
\begin{IEEEbiography}{Edyta Paulina Bogucka}{\,}is a Doctoral Candidate in Cartography at the Technical University of Munich, Germany. Her current research interests involve visual storytelling, spatial digital humanities, and urban cartography. Contact her at e.p.bogucka@tum.de.
\end{IEEEbiography}

\begin{IEEEbiography}{Marios Constantinides}{\,}is a Research Scientist in the Social Dynamics team at Nokia Bell Labs Cambridge (UK). He is interested in human-computer interaction, ubiquitous and affective computing. Contact him at marios.constantinides@nokia-bell-labs.com.
\end{IEEEbiography}

\begin{IEEEbiography}{Luca Maria Aiello}{\,}is a Senior Research Scientist in the Social Dynamics team at Nokia Bell Labs Cambridge (UK). He conducts interdisciplinary research in network science, computational social science, and urban informatics. Contact him at luca.aiello@nokia-bell-labs.com.
\end{IEEEbiography}

\begin{IEEEbiography}{Daniele Quercia}{\,}is the Department Head at Nokia Bell Labs in Cambridge (UK) and Professor of Urban Informatics at Kings College London. His research focuses in the areas of data mining, computational social science, and urban informatics. Contact him at quercia@cantab.net.
\end{IEEEbiography}

\begin{IEEEbiography}{Wonyoung So}{\,} is a Doctoral Candidate at the Massachusetts Institute of Technology (MIT). His is interested in the bottom-up data movement, and in how the open-source and DIY movement will affect the urban space. Contact him at mail@wonyoung.so.
\end{IEEEbiography}

\begin{IEEEbiography}{Melanie Bancilhon}{\,}is a Doctoral Candidate in Washington University St. Louis. She is interested in in human-computer interaction, urban informatics and computational social science. Contact her at mbancilhon@wustl.edu.
\end{IEEEbiography}